\documentstyle[11pt,newpasp,twoside,epsf]{article}
\def\edcomment#1{\iffalse\marginpar{\raggedright\sl#1\/}\else\relax\fi}
\reversemarginpar

\begin{document}
\title{Probing the Initial Conditions of Clustered Star Formation -- 
Large Scale On-the-Fly Mapping of Orion B at FCRAO}
 \author{Naomi A. Ridge}
\affil{FCRAO, 619 Lederle GRC, University of Massachusetts, 
Amherst, MA 01003, USA.}
\author{Edwin A. Bergin \& S. T. Megeath}
\affil{Harvard-Smithsonian CfA, 60 Garden St., Cambridge, MA 02138, USA.}

\begin{abstract}
In order to obtain a census of the pre-stellar and star-forming
molecular cores, we have begun an unbiased survey in CS and N$_2$H$^+$ of
the L1630 and L1641 molecular clouds.  The use of these two molecular
species enables us to quantify and disentangle the effects of
depletion often seen in CS observations of dense cores. The spectral
line data will provide essential kinematical information not present
in similar studies of the sub-millimeter dust-continuum, enabling us
to examine the overall core to core velocity dispersion and study the
effects of infall and outflows around known sub-mm and infra-red
sources. Here we present our initial observations of part of L1630,
taken during the commissioning phase of the FCRAO
On-the-Fly Mapping system.
\end{abstract}

\section{Introduction}
CS is an often used tracer of dense gas in molecular clouds (e.g.
Lada et al.\ 1991 (LEF); Lada et al.\ 1997).
These and similar studies have shown that the CS emission is not
uniformly distributed, but confined to dense cores which constitute
just a small fraction of the total cloud volume and mass.  LEF were
able to identify all previously known dense regions of star-formation
in their CS study of L1630, as well as finding many previously unknown
condensations due to their CS emission.  It has now become customary
to search for star-forming cores using CS.

However, recent chemical models (summarized by Bergin 2000) have shown
that CS is likely to be depleted due to freezing onto dust grains in
the densest regions of pre-star-forming cores and therefore may not be
the best tracer of dense star-forming gas.  An alternative molecule to
use as a tracer for dense gas is N$_2$H$^+$. N$_2$H$^+$ traces similar
excitation conditions to CS, but chemically responds differently to
variations in density. As a result, CS is more likely to trace
post-star-forming gas while N$_2$H$^+$ traces the pre-star-forming
cores.

We have therefore begun an unbiased study of two clustered star
forming regions, L1630 and L1641 in both CS and N$_2$H$^+$. By mapping
the regions in both lines (which can be observed simultaneously at
FCRAO) we hope to be able to identify both the youngest cores, which
have undergone collapse but not yet formed stars (traced by the
N$_2$H$^+$) and post-star-forming cores (traced by the CS), and look
for evidence of the freezing out of CS. The combination of the
optically thick CS and optically thin N$_2$H$^+$ will also enable us
to search for evidence for infall in the clusters. This essential
kinematical information complements existing (sub)-millimetre dust
continuum observations of these cluster-forming regions.  Here we
present initial maps of the region surrounding LBS\,23 and NGC\,2068.

\section{Observations}

The observations were made with the FCRAO 14m telescope during the
commisioning period of an On-the-Fly (OTF) mapping scheme in 2002
January.  OTF mapping with the 16-element SEQUOIA\footnote{SEQUOIA was
successfully upgraded to a 32-element dual-polarisation system in
March 2002, providing even faster mapping capabilities.} array
receiver provides fast, high quality and high-sensitivity imaging.
Using the newly commissioned Dual-IF correlator with 25MHz total
bandwidth and 1024 channels (providing an effective velocity
resolution of 0.08\,kms$^{-1}$), the two lines were observed
simultaneously.  Maps were obtained by scanning in the RA direction,
and an ``off'' source reference scan was obtained after every two
rows.  Each $10'\times10'$ submap required 3 repeats to build up the
required sensitivity. The area shown here therefore represents a total
of 18 hours of on-source observing. System temperatures during the
observations were around 200\,K.

Maps obtained by this method are not evenly sampled due to the
rotation of the sky with respect to the array, and a convolution and
regridding algorithm has to be applied to the data in order to obtain
spectra on a regularly sampled grid.

\section{Initial Results}

Figure 1a shows an overlay of N$_2$H$^+$ 1--0 integrated intensity on
a map of the CS 2--1 integrated intensity, in the region of LBS\,23
and NGC\,2068. These maps cover $\sim600$ square arcminutes and
represent a pilot study of a much larger project. The emission maps
shown here contain $\sim4000$ spectra at 25$''$ spacing (Nyquist
sampling), each with an rms sensitivity of $\la$0.1\,K per 25\,kHz
channel.

\begin{figure}
\plotfiddle{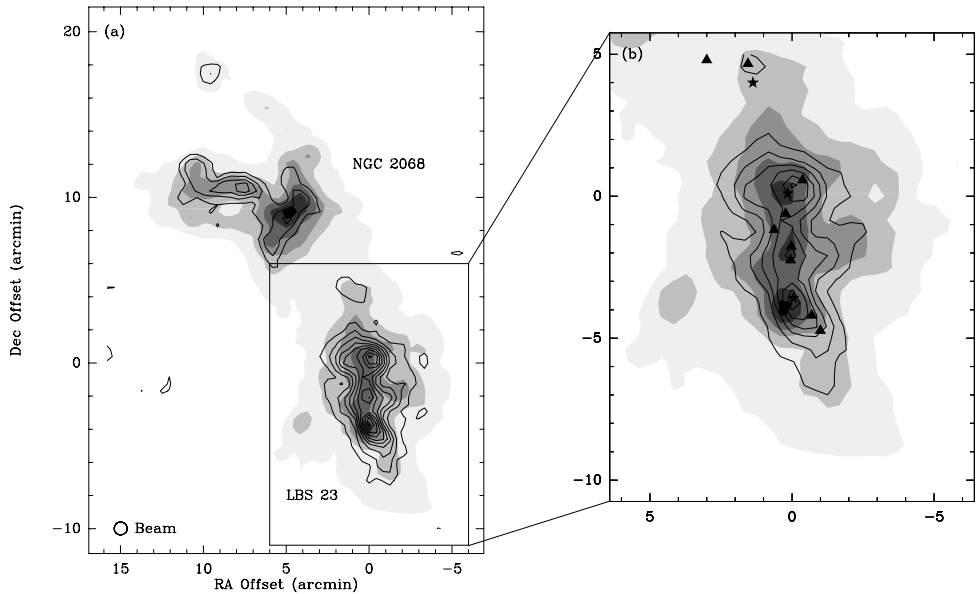}{2.78in}{0}{53}{53}{-195}{250}
\caption{Left: N$_2$H$^+$ contours overlaid on a grey-scale map of CS 2--1
emission.  Right: Enlarged map of the N$_2$H$^+$ (contours) and CS
(grey-scale) of the region around LBS\,23. The symbols indicate the
positions of sub-mm sources detected by Lis, Menten \& Zylka (1999).
The (0,0) position corresponds to the position of LBS\,23,
\hbox{$\alpha=$05:43:34}, \hbox{$\delta=-$00:11:12} (1950; LEF).}
\label{fig1}
\end{figure}

Our pilot study has already shown some intriguing results. The
N$_2$H$^+$ emission is less extended than the CS and a clear offset is
seen between the peaks of the N$_2$H$^+$ and CS emission. This offset
is not due to pointing error, as the data were obtained simultaneously
with the dual-IF system. Figure 1b shows the area around LBS\,23 in
more detail. Triangles indicate the positions of sub-mm continuum
sources found by Lis, Menten \& Zylka (1999) and the stars show the
positions of the sub-mm sources with IRAS detections. The sub-mm
sources form a chain along the bright CS ridge, but their positions
seem to correspond more closely to the N$_2$H$^+$ peaks, indicating
that N$_2$H$^+$ is a better tracer of the youngest star-forming cores
than CS.

The left panel of Figure 2 shows the CS map overlaid on a 2MASS
K-band image. The NGC\,2068 cluster is located to the north-east of
the CS core, in a region cleared of molecular gas as traced by CS.
Bright stars are seen in several regions of high CS column density.
The CS channel maps in the right panel of Figure 2 show structure not
visible in the integrated intensity. Four arched filaments stretch out
from the NGC\,2068 region, while the LBS\,23 region appears more
diffuse.  The northern region (around the NGC\,2068 cluster) is
present in more of the channels, indicating a higher CS line-width
than the more quiescent southern region.

\begin{figure}
\plotfiddle{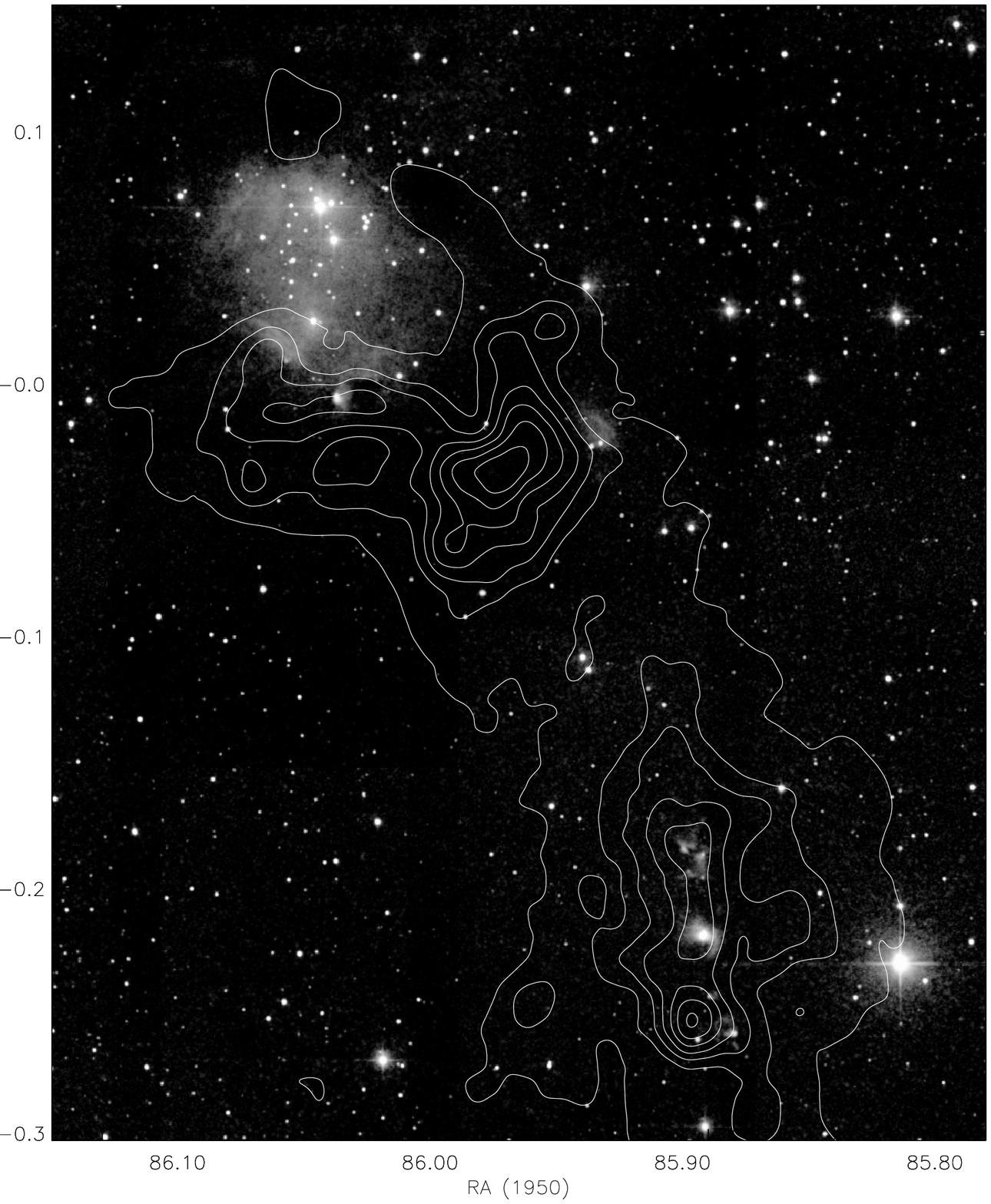}{3in}{0}{38}{38}{-190}{-30}
\plotfiddle{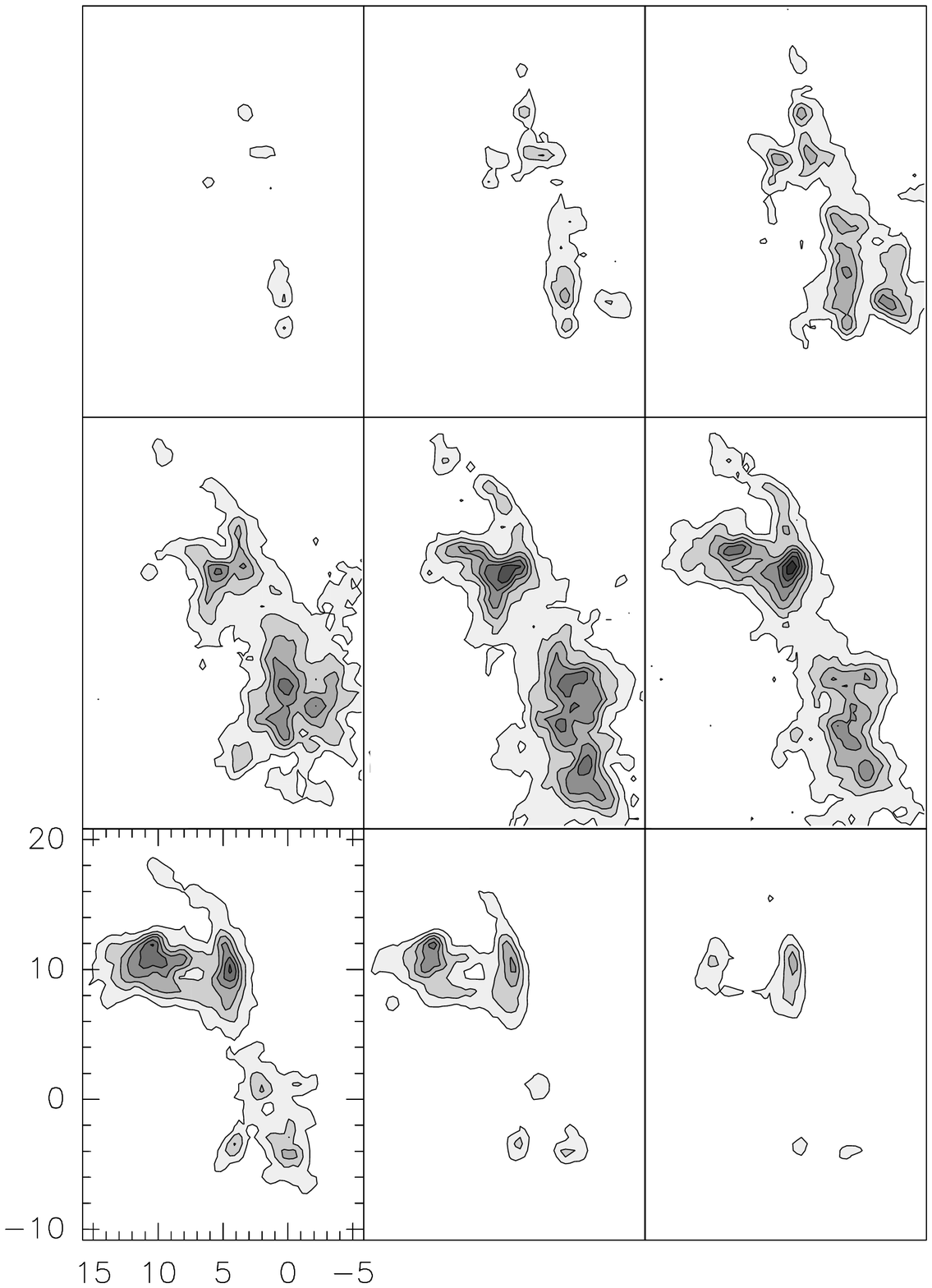}{0in}{0}{38.5}{38.5}{20}{-38}
\caption{Left: CS 2--1 contours overlaid on a 2MASS K-band image.
Right: 0.5\,km\,s$^{-1}$ channel-maps of CS integrated emission, from
8--12.5\,km\,s$^{-1}$\label{poo}}
\end{figure}

\section{Project Goals}

We are currently in the process of extending our survey to north to
include the NGC\,2071 region, and also to the Orion Nebula Cluster
region, L1641-South cluster, the NGC\,2023/4 region and the L1641
streamer.  This ambitious project, includes a range of star-forming
environments, from relatively quiescent to huge cluster-forming
regions.  In total our survey will cover approximately 9 sq.\ degrees.

The results from our pilot study already demonstrate the utility of
chemistry as a tool for star formation studies. Using only CS as a
tracer will bias a core sample to warmer or less dense regions (where
depletion is less evident). The combination of CS and N$_2$H$^+$ maps
probes a wider range of conditions.  L1630 and L1641 are known to
contain varied star-forming activity (Allen 1996), and therefore these
data will allow for a more complete comparison of the distribution of
molecular gas to known star-forming sites.

These two tracers are also routinely used for infall studies and CS is
a common tracer of outflows. The huge area of our survey will enable
us to look for these phenomena on scales previously beyond the
capability of millimetre observatories.

Our project goals are:

\begin{itemize}

\item To compare and
contrast core properties (mass, dynamics, densities) in the differing
environments of L1630 and L1641.

\item With the
use of 2MASS and existing (sub-)millimetre continuum data, to
investigate how core morphological and kinematical properties are
related to the YSO population, and look for spectral and chemical
signatures which trace different modes (clustered vs. distributed) of
star formation.

\item To investigate the effects of CS depletion, 
and look for an evolutionary trend in depletion between the cores, by
combining the CS and N$_2$H$^+$ data.

\end{itemize}

\acknowledgements
FCRAO is supported by NSF grant AST\,01--00793. The Two Micron All Sky
Survey is a joint project of the University of Massachusetts and the
Infrared Processing and Analysis Center/California Institute of
Technology, funded by the National Aeronautics and Space
Administration and the National Science Foundation

\end{document}